\documentclass[prl,aps,superscriptaddress,showpacs,amssymb,
amsmath,twocolumn]{revtex4}
\usepackage{bm,graphicx}

\newcommand{\edinburgh}{School of Physics, University of
Edinburgh, King's Buildings, Edinburgh EH9 3JZ, U.K.}

\newcommand{\glasgow}{Department of Physics and Astronomy, 
University of Glasgow, Glasgow G12 8QQ, U.K.}

\newcommand{\asqtad}{\textsc{asqtad}}
\newcommand{\hyp}{\textsc{hyp}}
\newcommand{\hisq}{\textsc{hisq}}

\newcommand{\tr}{\mathbin{\text{Tr}}}

\def\slashchar#1{\setbox0=\hbox{$#1$}           
   \dimen0=\wd0                                 
   \setbox1=\hbox{/} \dimen1=\wd1               
   \ifdim\dimen0>\dimen1                        
      \rlap{\hbox to \dimen0{\hfil/\hfil}}      
      #1                                        
   \else                                        
      \rlap{\hbox to \dimen1{\hfil$#1$\hfil}}   
      /                                         
   \fi}                                         %

\begin{document}

\title{The Index Theorem and Universality Properties of the 
Low-lying Eigenvalues of Improved Staggered Quarks}

\author {E. \surname{Follana}}
\affiliation{\glasgow}

\author{A. \surname{Hart}}
\affiliation{\edinburgh}

\author{\ ${}$C.T.H. \surname{Davies}}
\affiliation{\glasgow}
\collaboration{${}$HPQCD and UKQCD collaborations}

\begin{abstract}
  
  We study various improved staggered quark Dirac operators on
  quenched gluon backgrounds in lattice QCD generated using a
  Symanzik-improved gluon action. We find a clear separation of the
  spectrum into would-be zero modes and others. The number of would-be
  zero modes depends on the topological charge as expected from the
  Index Theorem, and their chirality expectation value is large
  ($\approx$ 0.7). The remaining modes have low chirality and show
  clear signs of clustering into quartets and approaching the random
  matrix theory predictions for all topological charge sectors. We
  conclude that improvement of the fermionic and gauge actions moves
  the staggered quarks closer to the continuum limit where they
  respond correctly to QCD topology.

\end{abstract}

\preprint{Edinburgh 2004-09}
\preprint{GUTPA/04/06/01}

\pacs{11.15.Ha, 
      12.38.Gc  
     }

\maketitle

\section{Introduction}
\label{sec_introduction}

It has been widely held that lattice staggered quarks are
insensitive to the topology of the underlying gauge fields. The
low-lying spectrum of the Dirac operator has neither shown the number
of chiral (near-) zero modes anticipated from the Index Theorem, nor
have the eigenvalues lain on the expected universal distributions.

Here we present evidence that this is not a generic failing of
staggered quarks, but simply a problem of discretisation errors, and
that the use of improved staggered Dirac operators clarifies the
situation and points to the correct continuum behaviour. 
This allows
the use of improved staggered quarks in lattice QCD to study
topologically sensitive states, such as those associated with the
axial anomaly (principally the $\eta^\prime$ meson). It also has
a bearing on establishing the effect of taking the fourth
root of the staggered determinant to represent one flavour of
staggered sea quarks.

We begin by reviewing our understanding in the continuum. The
eigenmodes of the (anti--Hermitian, gauge covariant) massless Dirac
operator are given by
\begin{equation}
\slashchar{D} f_s = i \lambda_s f_s \; , ~~~~~ 
\lambda_s \in \mathbb{R} \; .
\end{equation}
where we use orthonormalised eigenvectors, $f_s^\dagger f_t =
\delta_{st}$. As $\{ \slashchar{D},\gamma_5 \} = 0$, the spectrum is
symmetric about zero: if $\lambda_s \not = 0$, then $\gamma_5 f_s$ is
also an eigenvector with eigenvalue $-i \lambda_s$, and chirality
$\chi_s \equiv f_s^\dagger \gamma_5 f_s = 0$. The zero modes,
$\lambda_s = 0$, can be chosen with definite chirality: $\chi_s = \pm
1$. In general there are $n_{\pm}$ such modes, whose relative number
is fixed by the (gluonic) topological charge
\begin{equation}
Q = \frac{1}{32 \pi^2} \int d^4x \; \epsilon_{\mu \nu \sigma \tau}
\tr F_{\mu \nu}(x) F_{\sigma \tau}(x)
\end{equation}
via the Atiyah--Singer Index Theorem
\cite{Atiyah:1963,Atiyah:1968}:
\begin{equation}
Q = m \tr \frac{\gamma_5}{\slashchar{D} + m} = n_+ - n_- \; ,
\label{eqn_index}
\end{equation}
where $m$ is the quark mass. Based on 
\cite{Leutwyler:1992yt},
it has been suggested that (for a sufficiently large volume) the
non-zero low-lying eigenmodes
take values from a universal distribution
\cite{Shuryak:1993pi}
scaled by a QCD-specific quantity (the chiral condensate). 
The universality class is determined by the chiral symmetries of QCD
%
%
with separate predictions for each sector of fixed topological charge.
The distributions can be derived from any theory in the correct
universality class, such as ensembles of random matrices
%
\cite{Nishigaki:1998is,Damgaard:2000ah}
(for a review of other theories, see
\cite{Damgaard:2001ep}).

The four-dimensional lattice staggered massless Dirac operator is
anti-Hermitian, with a purely imaginary spectrum $i \lambda_s$.  It
represents $N_t = 4$ ``tastes'' of fermions that interact via
highly-virtual gluon exchange at finite lattice spacing, causing
taste-symmetry violations \cite{Lepage:1998vj}. These vanish in the
continuum limit (as $a^2$) and we then expect to recover a four-fold
degeneracy in the spectrum.  A remnant of continuum chiral symmetry in
the staggered action gives a local and taste-non-singlet $\gamma_5$
operator that guarantees that the spectrum is symmetric about zero, as
in the continuum. The $\gamma_5$ operator, $\gamma_5^{ts}$, relevant
to the index theorem must be a taste-singlet one, however, since only
this can couple to the vacuum correctly \cite{Golterman:1984}. As $\{
\slashchar{D},\gamma_5^{ts} \} \not = 0$, there is no exact Index
Theorem and all eigenmodes in principle contribute to
Eqn.~(\ref{eqn_index})
\cite{Smit:1987fn}.

If the gauge field is sufficiently close to the continuum limit,
however, we expect to see the continuum features developing. There
should be $2|Q|$ near-zero modes on either side of zero, whose
chiralities are close to unity. The taste-singlet $\gamma_5^{ts}$
operator is not conserved so we expect a renormalisation to achieve a
value of 1 in the continuum.  The remaining modes should have
chirality near zero, and come in approximately degenerate quartets on
either side of zero.  The values of the eigenvalue quartets should be
described by the same universal distribution as continuum QCD, up to a
renormalisation of the chiral condensate.

For thermalised lattices at finite lattice spacing, the fluctuations
in the gauge field can lead to a breakdown of this picture. This
happens for the simplest, ``one-link'' (na\"{\i}ve) staggered operator
with the unimproved, Wilson gauge action at values of the gauge
coupling used in present-day simulations.  The breakdown is seen in
two ways.  Firstly, there is no clear separation (in eigenvalue or
chirality) of near-zero modes of topological origin from the remainder
of the spectrum
\cite{Smit:1987fn,Hands:1990,Venkataraman:1998yj,Hasenfratz:2003,Follana:2003}.
In addition, the low-lying eigenvalues do not follow the predictions
from universality. In fact, the eigenvalues in each sector of charge
$Q \not= 0$ all follow the distribution corresponding to the sector with
$Q=0$
\cite{Berbenni-Bitsch:1998tx,Damgaard:1998ie,
  Gockeler:1998jj,Damgaard:1999bq,Damgaard:2000qt}.  
(It should be noted that the method we follow here, of grouping the
eigenvalues into quartets, was not followed because this feature of
the spectrum was not evident.)

This failure can be ascribed to taste-changing interactions and the
lack of a good continuum chiral symmetry in the one-link staggered Dirac
operator: good agreement with predictions for all topological charge
sectors has been seen for Dirac operators obeying the Ginsparg-Wilson
relation
\cite{Edwards:1999ra,Edwards:1999zm,Damgaard:1999tk,
Hasenfratz:2002rp,Bietenholz:2003mi,Giusti:2003gf}.

Over the past few years considerable advances have been made in
lattice QCD phenomenology through the use of so--called ``improved
staggered'' fermion formulations
\cite{Davies:2003ik}.
The goal of this programme is to systematically reduce the lattice artefact taste
interactions, and it is reasonable to expect that this will improve
the continuum-like chiral properties of the fermions.  The variation
of the topological susceptibility with the sea quark mass, in
particular, shows that this is so: whilst it was largely insensitive
to the presence of one-link staggered sea quarks
\cite{Bitar:1991wr,Kuramashi:1993mv,Alles:2000cg,Hasenfratz:2001wd},
the improved operator gives a variation with $m$ that agrees well with
theoretical expectations
\cite{Hasenfratz:2001wd,Bernard:2003gq}.
It is thus pertinent to study in more detail the spectrum of the
improved staggered Dirac operator, and we shall show here that this
also now gives signs of converging to the expected results, in
contrast to previous studies.
\begin{figure}
\includegraphics[width=3in,clip]{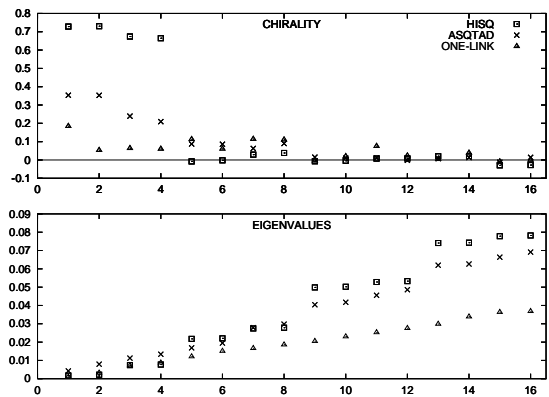}

\caption{\label{fig_spectrum} The positive half of a
  typical low-lying eigenmode spectrum for a configuration of $Q = 2$,
  for various staggered fermion formulations.  The bottom panel gives
  the absolute value of the eigenvalue, $\lambda_s$, ordered according
  to increasing size. The $x$ axis is then simply eigenvalue number.
  The top panel is the chirality of the modes. The \hyp\ action gives
  results very similar to \hisq\ and is not plotted for clarity.}
\end{figure}
\section{Results}

On a Euclidean lattice with lattice spacing $a$, the one-link massless
staggered operator is
\begin{equation}
\slashchar{D}(x,y) = \frac{1}{2a} \sum_{\mu=1}^4
\eta_\mu(x) \left[ U_\mu(x) \delta_{x+\hat{\mu},y} - H.c. \right] \; ,
\label{eqn_naive}
\end{equation}
with $\eta_1 = 1$ and $\eta_\nu = (-1)^{x_\nu} \eta_{\nu-1}$. 
We study also three improved operators: the \asqtad\
\cite{Lepage:1998vj,Blum:1997uf,Bernard:1999xx,Orginos:1999cr},
the \hyp\
\cite{Knechtli:2000ku}
and the Highly Improved Staggered Quark (\hisq)
\cite{Follana:2003,Follana:2004}.
These operators use ``smeared'' gauge fields in place of the $U$ field
above, obtained by multiplying $U$ fields along combinations of bent
paths from the start to end points of the original link.  This reduces
the coupling to highly virtual gluons and suppresses the
taste-changing interactions. The \asqtad\ action uses a ``FAT7''
smearing, which includes paths made of up to seven links, and tadpole
improvement. The \hyp\ action uses an hypercubic blocking procedure,
involving reunitarization back onto SU(3), and the \hisq\ action uses
two applications of the FAT7 smearing, and also includes
reunitarization. The taste-changing interactions are most suppressed
for the \hyp\ and \hisq\ cases. The \hisq\ and \asqtad\ cases are
completely $a^2$ improved at leading order by additional terms which
correct for errors in the simple derivative above.

We calculate the eigenmodes for these Dirac operators on an ensemble
of quenched (no sea quarks) SU(3) gauge configurations. The gauge
action is Symanzik-improved at tree-level with tadpole
improvement so that remaining discretisation errors from the gluon field are a
small number times $\alpha_s a^2$.  The lattice spacing is $0.093$~fm,
representing a standard ensemble in present-day lattice simulations
\cite{Zhang:2001fk}.
The majority of our results are from 1000 configs on a periodic lattice with $16^4$ sites,
which should be large enough that finite volume effects on the
low-lying spectrum are negligible
\cite{Giusti:2003gf}.
We have also studied a larger volume, $24^4$. On each configuration we
determine the topological charge by two standard methods that involve
cooling the gauge fields 
\cite{Bernard:2003gq}.
The calculation of gluonic topological
charge always involves ambiguities and we discard those configurations
(10\% of the ensemble) for which the two cooling methods do not agree,
to leave a sample for which we are confident that we have a robust
estimate of $Q$. We stress at this point that cooling is used
\textit{solely} to determine the gluonic topological charge, and that
all the Dirac eigenmode calculations are carried out on fully
thermalised, ``raw'' configurations.

Fig.~\ref{fig_spectrum} compares the low-lying modes of the spectra of
the various staggered quark formulations on a typical background with
$Q=2$. We show the eigenvalue $\lambda_s$ and chirality for the upper
half of the spectrum. We see that the Index Theorem is well
approximated for the more improved quark formulation. Specifically,
there is a clear delineation between the near-zero modes (small
eigenvalues and large chirality) and the rest of the spectrum.  The
number of near-zero modes is $2|Q|$, as expected. The other
eigenvalues have small chirality, and divide clearly into quartets. The
taste-singlet $\gamma_5^{ts}$ operator used for the chirality is a
point-split 4-link operator. We measure it here by inserting $U$
fields to make it gauge-invariant. We will report elsewhere on the
dependence of the chirality on the operator used.

As the near-zero modes for the more improved actions clearly separate and
have well defined chirality, we may define an index $\bar{Q} = n_+ -
n_-$ on each configuration. $\bar{Q}$ is then strongly correlated with $Q$. 
If we count eigenmodes with absolute value of the
chirality above 0.65 in $n_{\pm}$, for example, we find that in about 90 \% of
configurations $\bar{Q}$ and $Q$ are the same, satisfying the Index Theorem. 
Indeed, measuring $Q$ from the Index Theorem then becomes as reliable 
as measuring it from gluonic methods.  

\begin{figure}
\includegraphics[width=3in,clip]{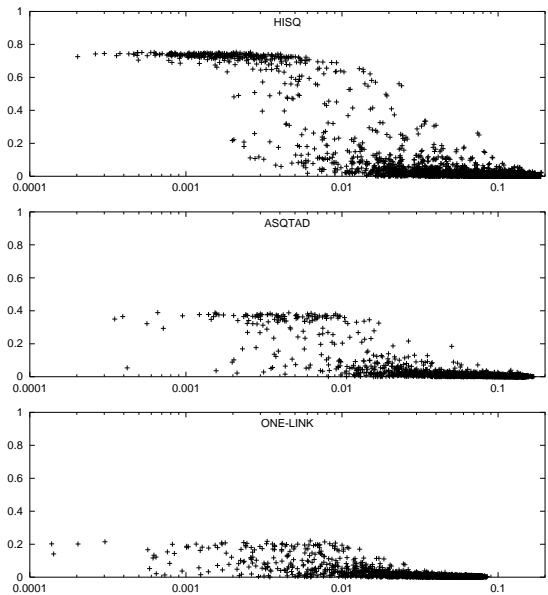}

\caption{\label{fig_scatter} A scatter plot for different staggered quark formulations, with absolute value of the
chirality on the y axis and eigenvalue $\lambda_s$ on the x axis. The lowest 
50 eigenvalues for 147 configurations are plotted. }

\end{figure}

In Fig.~\ref{fig_scatter} we show a scatter plot of the absolute value
of the chirality versus the (absolute value of) eigenvalues, for
different operators. We can see the formation of a gap between modes 
of small and large chirality 
as we improve the staggered operators, as well as the
overall increase in chirality of the large chirality modes. 

\begin{figure}
\includegraphics[width=3in,clip]{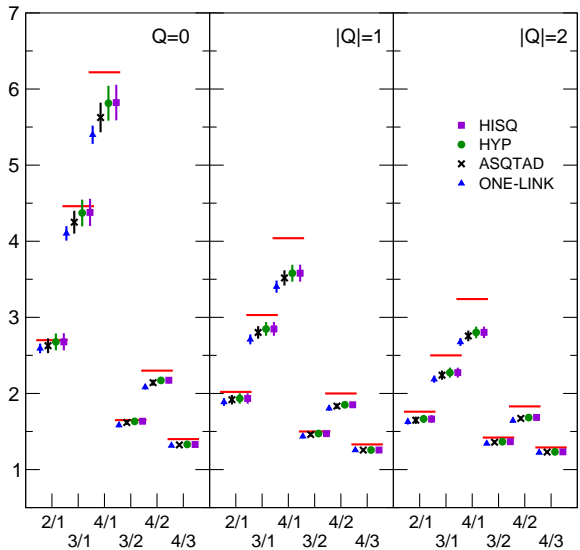}

\caption{\label{fig_mean_eig} The ratios of expectation values of
  small eigenvalues (see text for notation) compared with the
  predictions based on a universal distribution (horizontal lines) for topological charge
  sectors 0, 1 and 2.}

\end{figure}

We turn now to the non--zero modes. Subtracting the $2 |Q|$ near-zero
modes from the spectrum, we group the other eigenvalues, ordered by
size, into sets of four, as indicated in the \hisq\ and \hyp\ cases in
Fig.~\ref{fig_spectrum}. We call the average of these sets
$\Lambda_{1,2,\dots}$. In Fig.~\ref{fig_mean_eig} we plot the ratios
$\langle \Lambda_s \rangle_Q / \langle \Lambda_t \rangle_Q$ (denoted
by ``$s/t$''), where the expectation values $\langle \cdot \rangle_Q$
are over configurations with gluonic topological charges $\pm Q$ only.
Also shown are the universal predictions. There is a clear dependence
of the ratios on $Q$, in marked contrast with previous results, which
showed a precise agreement with the $Q = 0$ predictions for all
sectors.

The results are systematically slightly lower than the theoretical predictions,
especially for the ratios involving higher eigenvalues. This would be
consistent with finite volume effects as in
\cite{Giusti:2003gf}. 
There is also a small but systematic difference between the one-link
and the improved actions, with the improved results showing a better
agreement with the theoretical values.

An important point to make here is that it is necessary to group the
eigenvalues as explained above to get sensible results. If one ignores
the near zero modes, or does not group in quartets, ratios which are
close to one or very large will result.  This is strong evidence that
the four tastes are showing up in the spectrum, 
even where it is not directly evident in the spectrum itself.
We are undertaking further analysis to understand why previous results 
with naive staggered quarks on unimproved gluon fields were in agreement with 
universal distributions for $Q$ = 0 for all $Q$ values when this procedure was 
not followed. 
It seems clear from our 
analysis here that results on unimproved gluon fields with a fine 
enough lattice spacing should show $Q$-dependent universal distributions 
when zero modes and quartets are taken into account. Preliminary results 
confirm this expectation. 
\section{Conclusions and Outlook}

Improved staggered fermions are not blind to the topology, but in fact
reproduce well the predictions of the Index Theorem, and the
universality of ratios of eigenvalues as a function of topological
sector.  This means we can have confidence in using them to attack the
questions arising from the axial anomaly in QCD.

We also remark that the fact that the 4-fold taste degeneracy of staggered 
quarks is becoming clear in the spectrum is encouraging for the programme 
of establishing the effect of taking the fourth root of the staggered 
determinant to represent one flavor of staggered sea quarks. This programme 
requires an analysis in the taste basis and progress towards this is now 
possible. 

More extensive studies of finite volume and lattice spacing effects
and analysis of the eigenvectors are underway and will be reported
elsewhere.

In the later stages of this study we became aware of work 
on a related topic
\cite{Wenger}.
\section{acknowledgements}
  
  We thank: Ph. de Forcrand for his topological charge
  measurement code;  A. Hasenfratz for help in implementing
  the \hyp\ operator; and P. Lepage for many useful discussions.
  E.F. and C.D. are supported by PPARC and  A.H. by the
  U.K. Royal Society.  The eigenvalue calculations were carried out on
  computer clusters at Scotgrid and the Dallas Southern
  Methodist University. We thank David Martin and Kent Hornbostel for
  assistance.

\bibliographystyle{h-physrev4}
\bibliography{rmt_refs}

\end{document}